\documentclass[11pt]{amsart}
\pdfoutput=1
\usepackage{times}
\usepackage{comment}
\usepackage{epsfig}
\usepackage{varioref}
\usepackage{textcomp}
\usepackage{multirow}
\usepackage{graphics}
\usepackage{bmpsize}
\usepackage{url}
\usepackage{draftcopy}
\usepackage[numbers]{natbib}
\usepackage{amssymb}
\usepackage{algorithm}
\usepackage{algpseudocode}
\usepackage{varioref}
\usepackage{multirow}
\usepackage{enumitem}
\makeatletter\@ifclassloaded{beamer}{}{\usepackage{imakeidx}}\makeatother
\usepackage{lscape}
\usepackage{gensymb}
\usepackage[toc]{appendix}
\usepackage{mathtools}
\usepackage{pict2e}
\usepackage{picture}

\makeatletter\@ifclassloaded{beamer}{}{}\makeatother

\newcommand{\genlab}[2]{\label{#1:#2}}
\newcommand{\genref}[2]{#1~\vref{#1:#2}}

\newcommand{\seclab}[1]{\genlab{section}{#1}}
\newcommand{\secref}[1]{\genref{section}{#1}}

\newcommand{\eqnlab}[1]{\genlab{equation}{#1}}
\newcommand{\eqnref}[1]{\genref{equation}{#1}}

\newcommand{\tablab}[1]{\genlab{table}{#1}}
\newcommand{\tabref}[1]{\genref{table}{#1}}

\newcommand{\figlab}[1]{\genlab{figure}{#1}}
\newcommand{\figref}[1]{\genref{figure}{#1}}

\makeatletter\@ifclassloaded{beamer}{}{%

}
\makeatother
\newcommand{\pdffigure}[4]{\begin{center}\begin{figure}[hbt]\includegraphics[angle=#4,scale=#3]{#1.pdf}\caption[\hspace{\normalparindent}#2]{#2}\figlab{#1}\end{figure}\end{center}}
\makeatletter\@ifclassloaded{beamer}{\input{beamer}}{}\makeatother

\usepackage{booktabs}
\vrefwarning
\begin{document}
%
%
\title[Correlation without Factors]{Correlation Without Factors in Retail Cryptocurrency Markets}
\author{Graham L. Giller}
\email{graham@gillerinvestments.com}
\date{\today}
\begin{abstract}
A simple model-free and distribution-free statistic, the functional relationship between the number of ``effective'' degrees of freedom and portfolio size or $N^*(N)$, is used to discriminate between two alternative models for the correlation of daily cryptocurrency returns within a retail universe of defined by the list of tradable assets available to account holders at the \textit{Robinhood} brokerage. The average pairwise correlation between daily cryptocurrency returns is found to be high (of order 60\%) and the data collected supports description of the cross-section of returns by a simple isotropic correlation model distinct from a decomposition into a linear factor model with additive noise with high confidence. This description appears to be relatively stable through time.
\end{abstract}
\maketitle
\section{Introduction}
The emergence of cryptocurrencies (see ``Nakamoto''\cite{nakamoto2008bitcoin}, Buterin\cite{buterin2014next} et al.) as an asset class available to traders and investors has prompted researchers and practitioners in quantitative finance to seek to deploy the factor model frameworks developed using the \textit{Arbitrage Portfolio Theory} of Ross\cite{ross2013arbitrage}, and that have gained widespread adoption in equity markets (see Asness\cite{asness2013value}, Muller\cite{muller1999private,muller2001proprietary} and many others), to the returns of this new class of synthetic commodities. It is readily apparent that the intertemporal returns of many different cryptocurrencies exhibit strong correlation cross-sectionally. This has lead to a series of studies in which methods such as \textit{Principal Components Analysis}\cite{mardia1979} and \textit{Factor Analysis}\cite{harmon1976modern} have been applied to data over a variety of horizons. Much of this work, such as that in Li\cite{li2018toward}, Liu\cite{liu2019factor}, Peng\cite{peng2024composite} and others, reproduces the approach pioneered by Fama and French\cite{fama1992cross}, constructing ``factors'' as the returns of portfolios composed by the author of the study by stratifying assets according to some \textit{ex ante} property, such as ``size'' or ``book value'' etc.\ in the equity world, and forming a panel regression of asset returns on to these synthetic time-series. These panel regressions extract the \textit{risk premium} attributable to each factor via the Fama-MacBeth two stage regression procedure\cite{fama1973risk}, and has given rise to the concept of \textit{Factor Investing} which is implemented by active managers throughout the finance industry\cite{bender2013foundations}. 

It is such a common practice for authors to invent a new, ad hoc, asset stratification rule and demonstrate that returns regress onto it without much consideration as to whether there is sufficient available residual variance to be explained by this new ``factor,'' or whether it is statistically independent of other known factors as it should be under the proper definition of a factor model, that the confusion arising from all of this work has generated a situation known to those who study the cross-section of equity returns as ``the Factor Zoo.''\cite{harvey2019census} 

In this article a different approach to understanding the cross-section of cryptocurrency returns is taken. Following the method developed in Giller\cite{giller2024isotropic}, a systematic study of the manner in which the variance of daily returns accumulates in equal-weighted portfolios of cryptocurrencies is made via random sampling. This produces a signature function, $N^*(N)$, which represents the ``effective'' degrees of freedom, $N^*$, as a function of the number of assets in the portfolio, $N$. This function is known to have very specific forms for both linear factor models, such as the A.P.T.\ models of Ross discussed above, and for a simpler isotropic covariance model presented by authors such as Grinold and Kahn\cite{grinold2000active}, mostly as a didactic tool. These forms differ and it will be shown that the data extracted from cryptocurrency markets favours strongly the latter over the former.

\section{Isotropic Returns and Linear Factor Models}\seclab{toymodel}
\subsection{Effective Degrees of Freedom in the Cross-Section of Asset Returns}
The concept of ``effective'' degrees of freedom is extremely straightforward to understand and very simple to compute. It is well known from the \textit{Law of Large Numbers}\cite{kendall1999vol1} that the variance of a mean of i.i.d.\ random variables, $\overline{x}$, is simply given by $\sigma^2/N$, where $\sigma^2$ is their common individual variance and $N$ is the sample size. If, alternatively, the variables have different variances, $\{\sigma_i^2\}$ for $i\in[1,N]$, while still being independent, then 
\begin{equation}\eqnlab{lln}
    \mathbb{V}[\overline{x}]=\frac{\overline{\sigma^2}}{N},
\end{equation}
where $\overline{\sigma^2}$ is the mean of the variances of the variables. In the case of positive (negative) correlation between the variables the variance of the mean will be more (less) than that expected under the false assumption that the variables were independent. This can be expressed in terms of an adjusted expression
\begin{equation}
    \mathbb{V}[\overline{x}]=\frac{\overline{\sigma^2}}{N^*},
\end{equation}
where $N^*$ is the ``effective'' degrees of freedom. 

The precise form of $N^*(N)$ depends on the specific correlation matrix that describes the variables. However, when the variables are the returns of assets that are collected into an equal-weighted portfolio, it is simple to show that
\begin{equation}
    N^*=N\frac{V_I}{V_P}.
\end{equation}
Here $V_I$ represents the variance expected for the portfolio if the assets were independent, which may be computed from the individual sample variances of the asset returns, and $V_P$ represents the actual sample variance of an equal-weighted portfolio formed from the assets.
\section{Measuring the Effective Degrees of Freedom}
\subsection{Experimental Design}\seclab{experiment}
An \textit{elementary} random sampling experiment may be constructed to investigate the form of the $N^*(N)$ function for a given set of assets. Importantly, this experiment is entirely \textit{independent} of any specific models for asset returns and so its results are not biased due to mis-specification errors.

For a given set of $N_\textrm{max}$ assets, $\mathcal{U}$, and sample period, $\mathcal{T}=[1,T]$, the following is performed repeatedly:
\begin{enumerate}
    \item set the iteration to $j\in[1,N_\mathrm{iter}]$, for $N_\mathrm{iter}$ total iterations;
    \item choose $N_j$ uniformly from $[1,N_\textrm{max}]$;
    \item choose a unique subset $\mathcal{S}_j$ of $N_j$ assets uniformly from within the available universe, $\mathcal{S}_j\subseteq\mathcal{U}$ where $|\mathcal{S}_j|=N_j$;
    \item from the sets of returns selected, $\{r_{it}\}:i\in\mathcal{S},\,t\in\mathcal{T}$, compute the equal-weighted portfolio variance, $V_P$, and the portfolio variance expected for independence, $V_I$;
    \item compute $N^*(N_j)=N_j\times V_I/V_P$;
    \item repeat until a sufficiently large sample of data is collected.
\end{enumerate}

The collected data may then be analyzed to compare it with various hypotheses regarding the structure of the cross-section of returns in a manner independent of any distributional assumptions. For $N_\textrm{max}$ assets there are $2^{N_\textrm{max}}-1$ ways of picking a portfolio with between $1$ and $N_\textrm{max}$ assets in it. This can rapidly become a very large number and, due to the random sampling nature of the experiment described, it is not necessary to exhaustively enumerate every possible portfolio. i.e. The set of portfolios studied, $\{\mathcal{S}_j\}$, may be (much) smaller than the power set $2^\mathcal{U}$ and statistical inference may be applied to this data.\footnote{Technically the empty-set is excluded so $\{S_j\}\subseteq(2^\mathcal{U}\setminus\emptyset)$.}
\subsection{Independent Returns}
For fully independent returns
\begin{equation}\eqnlab{indepnstarn}
    N^*(N)=N.
\end{equation}
It is unlikely that this is the outcome of the experiment.\footnote{Based on the author's experience in financial markets.}
\subsection{Isotropic Returns}
For homoskedastic isotropic covariance, the covariance matrix of returns may be written\footnote{The notations $\mathbb{E}[\boldsymbol{x}]$ and $\mathbb{V}[\boldsymbol{x}]$ is used to denote the expectation and covariance matrix of $\boldsymbol{x}$, respectively.}
\begin{equation}\eqnlab{isocov}
    \mathbb{V}[\boldsymbol{r}_{t}]=\sigma^2\begin{pmatrix}
1&\rho&\cdots&\rho\\
\rho&1&\cdots&\rho\\
\vdots&&\ddots&\vdots\\
\rho&\rho&\cdots&1
\end{pmatrix},
\end{equation}
where $\boldsymbol{r}_t$ is a vector of asset returns for $N_\mathrm{max}$ assets over the period $(s,t]$. It is easy to generalize this model to the case of heteroskedastic returns by removing the scalar $\sigma^2$ and, instead, multiplying the given correlation matrix, on the left and on the right, by a diagonal matrix with the individual volatility of each asset along the diagonal. This generates a system suitable for use in the ``Dynamic Conditional Correlation,'' or DCC, models of Engle and Sheppard\cite{engle2001theoretical}.

For the homoskedastic structure of \eqnref{isocov}
\begin{equation}\eqnlab{isonstarn}
    N^*(N)=\frac{N}{1+(N-1)\rho}.
\end{equation}
If this model is true it is necessary that $\rho\in[1/(1-N),1]$ and\footnote{The former condition is derived in Giller\cite{giller2024isotropic} and the latter limit in Grinold and Kahn\cite{grinold2000active}.}
\begin{equation}
        \lim_{N\to\infty}N^*(N)=\frac{1}{\rho}.
\end{equation}
\subsection{Linear Factor Models}
A linear $K$-factor model for returns is the structure
\begin{align}    \boldsymbol{r}_t&=\boldsymbol{\mu}+B\boldsymbol{f}_t+\boldsymbol{\varepsilon}_t\\
    \mathrm{where}\ \mathbb{E}[\boldsymbol{f}_t]&=\boldsymbol{0}_K,\; \mathbb{E}[\boldsymbol{\varepsilon}_t]=\boldsymbol{0}_N\\ \mathrm{and}\;\mathbb{V}[\boldsymbol{f}]&=I_K,
    \mathbb{V}[\boldsymbol{\varepsilon}]=(\mathop{\mathrm{diag}}\boldsymbol{s})^2,\;
    \mathbb{V}[\boldsymbol{f},\boldsymbol{\varepsilon}]=0\\
    \Rightarrow\ \mathbb{E}[\boldsymbol{r}_t]&=\boldsymbol{\mu}\;
    \mathrm{and}\ \mathbb{V}[\boldsymbol{r}_t]=B\,\mathbb{V}[\boldsymbol{f}]B^T+(\mathop{\mathrm{diag}}\boldsymbol{s})^2.
\end{align}
With this cross-sectional model, the effective degrees of freedom is given by\footnote{Ibid.}
\begin{equation}\eqnlab{factornstarn}
    N^*(N)=N\frac{\overline{b^2}N+\overline{s^2}}{\overline{\boldsymbol{b}}^T\overline{\boldsymbol{b}}N+\overline{s^2}}
\end{equation}
where $\overline{b^2}$ is the mean-squared element of the factor loadings matrix, $B$, and $\overline{\boldsymbol{b}}$ is the mean factor loadings vector. $\overline{s^2}=\boldsymbol{s}^T\boldsymbol{s}/N$ is the mean residual variance. All of the terms in \eqnref{factornstarn} are non-negative.\footnote{In addition, $N$ is a positive integer.}
\subsection{Discrimination Between Covariance Models}
It can be seen that equations \ref{equation:indepnstarn}, \ref{equation:isonstarn} and \ref{equation:factornstarn} exhibit \textit{different} functional forms for the measure $N^*(N)$ and that, to discriminate fully between them, it is important to explore the shapes of the functions not merely the specific values for a given portfolio size.

In particular, the asymptotic properties of the expressions are quite different: for isotropic correlation the function converges to single value, $1/\rho$, whereas for both independence and factor models it diverges linearly as $N$. For a model containing $K$ independent factors the coefficient of $N$ is scaled by $\overline{b^2}/\overline{\boldsymbol{b}}^T\overline{\boldsymbol{b}}$ which may be approximated by $1/K$ when the elements of the factor loading matrix, $B$, are all similar in value. However, such asymptotics are not possible in the analysis presented here because the total number of assets studied (see \secref{empirical}) is low. However, such an analysis is feasible for equity indices, such as the \textsc{S\& P~500} or \textsc{Russell~3000}.
\section{Analysis of Data Collected from Robinhood}\seclab{empirical}
Robinhood Markets, Inc. (hereafter ``Robinhood'') is a publicly traded retail brokerage.  Robinhood also operates a cryptocurrency trading platform for its clients in the U.S.A., under a subsidiary Robinhood Crypto, LLC. This platform offers a REST\cite{masse2011rest} API for clients to use, which is documented online\cite{robinhood2024api}.\footnote{The author holds several accounts at Robinhood and applied for access to the API as any other client would. See the appendix for fuller disclosures and descriptions of data and code access.}
\pdffigure{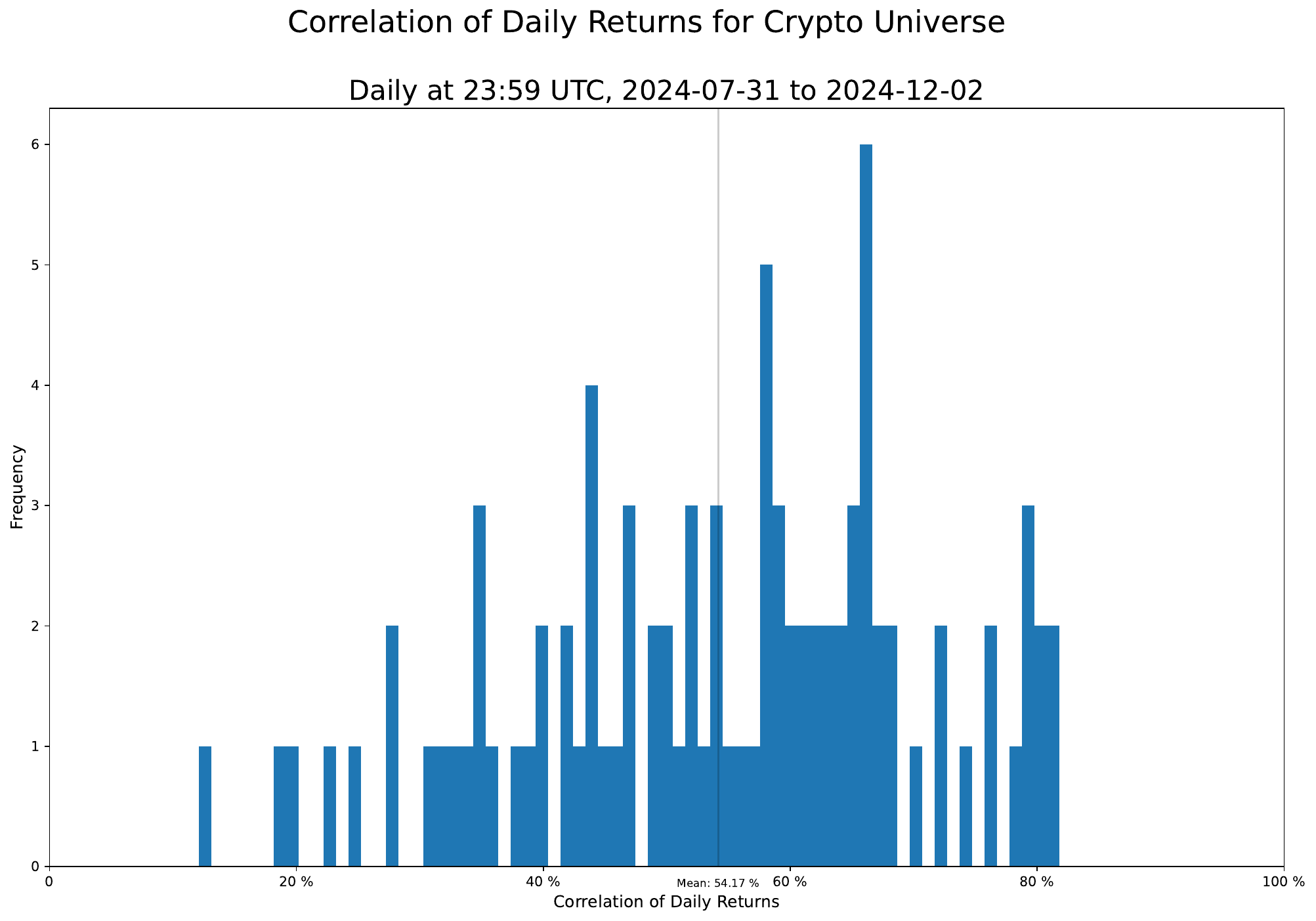}{Distribution of all possible pairwise correlation coefficients, $\rho_{ij}$, for the daily returns (UTC midnight to UTC midnight) of cryptocurrency prices collected from the \textit{Robinhood} retail brokerage.}{0.35}{0}
\subsection{Data Collection}
At the time of writing, Robinhood provided retail clients trading access to fourteen cryptocurrencies via their API. These are: \textsc{AAVE}\cite{frangella2022aave}, \textsc{AVAX}\cite{buttolph2020avax}, 
\textsc{BCH},\footnote{``Bitcoin Cash'': a fork of the original bitcoin blockchain with a modified block size.} \textsc{BTC}\cite{nakamoto2008bitcoin}, \textsc{COMP}\cite{leshner2019compound}, \textsc{DOGE}\cite{markus2013doge},  
\textsc{ETC}\footnote{``Ethereum Classic'': a copy of the original ethereum blockchain retaining the ``proof-of-work'' validation scheme.},
\textsc{ETH}\cite{buterin2014next}, 
\textsc{LINK}\cite{breidenbach2021link}, 
\textsc{LTC}\cite{lee2011ltc}, 
\textsc{SHIB}\cite{ryoshi2020shib}, 
\textsc{UNI}\cite{adams2018uniswap}, 
\textsc{XLM}\cite{mazieres2016stellar}, 
\textsc{XTZ}\cite{breitman2018tezos}. 
Data has been collected by sampling the best bid and ask quotes offered to Robinhood clients, every ten minutes, for every day, since the end of July, 2024. From this mid-price returns are computed. Data is naturally timestamped in Coordinated Universal Time (UTC) and all dates and times in this article are quoted in UTC. Only daily data will be analyzed herein, which is represented by the last print collected at 23:59 UTC on every date sampled. This data collection is ongoing and the data is available at the URL given by the reference cited in the appendix. Robinhood provides price data for a wider range of cryptocurrencies on its website, but these are not available within the API and analysis here is restricted to those that are tradable via the API by clients.
\subsection{Distribution of Pairwise Correlation}
Prior to executing the main experiment, as described in \secref{experiment}, a study of the pairwise correlations between daily returns of the selected universe of cryptocurrencies was made. As there are just $14\times13/2=91$ possible correlation coefficients to sample, the values for the entire population may be enumerated. This data is presented as a histogram in \figref{rho}. Qualitatively, the data shows a left-skewed distribution with a mean of 54.17\%. There are only three pairs of cryptocurrencies with correlations below 20\% and four above 80\%. With a sample size of 125 daily returns, the standard error of these correlation coefficients should be of order $1/\sqrt{122}\approx9\%$ (using Fisher's approximation\cite{fisher1915frequency}).
\pdffigure{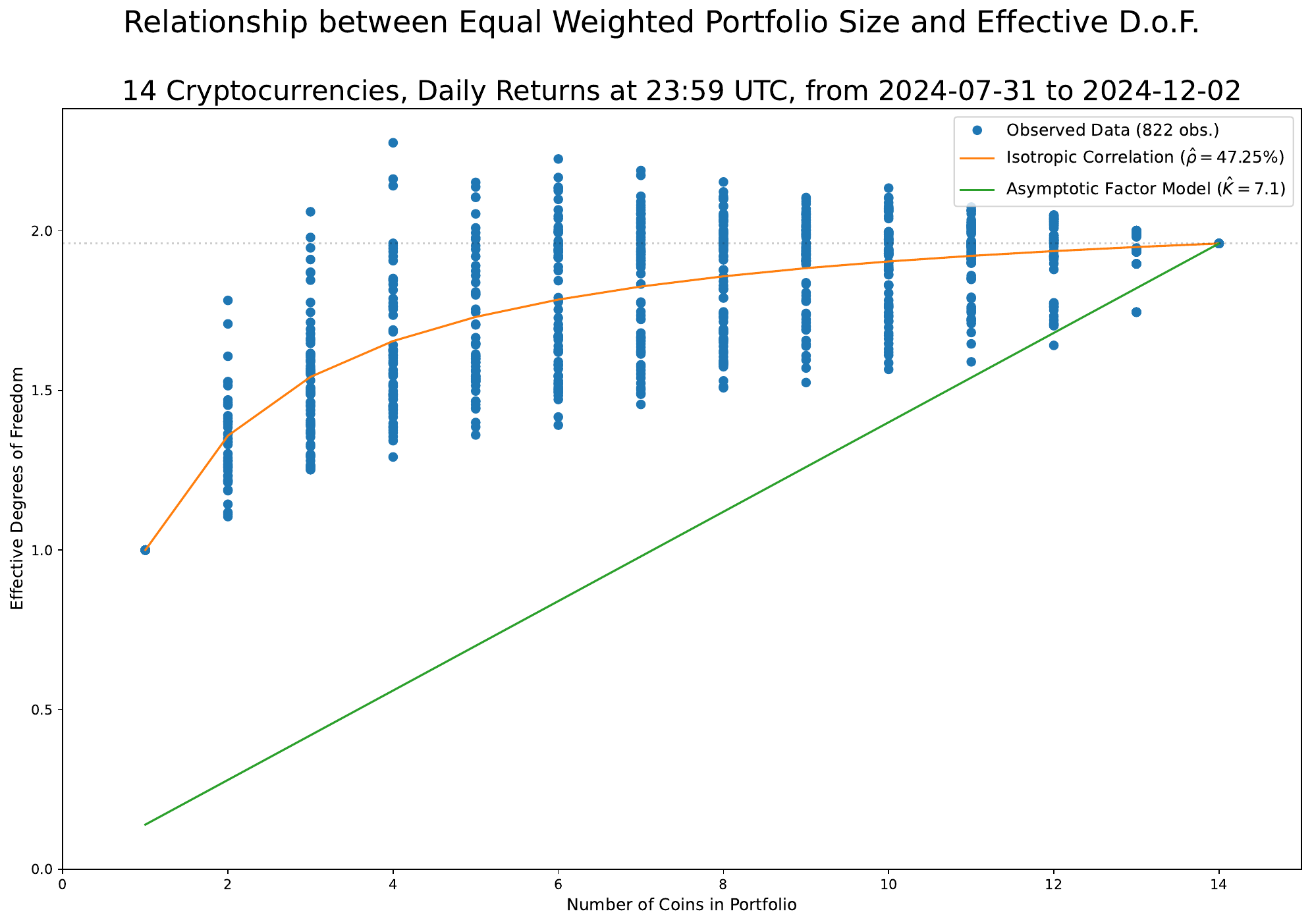}{A random sample of values (blue dots) of $N^*(N)$ for equal-weighted portfolios formed from the daily returns (UTC midnight to UTC midnight) of cryptocurrency prices collected from the \textit{Robinhood} retail brokerage. The orange curve represents the expected relationship for an isotropic correlation structure and the green line the asymptotic relationship expected for a linear factor model in a ``large'' portfolio.}{0.35}{0}
\subsection{Scaling of Effective Degrees of Freedom with Portfolio Size}
One thousand trials of the experiement described above were performed. For a universe size of $N_\mathrm{max}=14$ there are $2^{N_\mathrm{max}}-1$ ways of composing a portfolio with between one and fourteen assets. This is 16,383 and, in a thousand trials, the probability of picking the same asset mix more than once is non-negligible. To correct for this, the list of assets used for each portfolio is retained in the analysis and any duplicates are removed. The result is 822 distinct portfolio pairs of $(N^*,N)$, which are plotted in \figref{nstarn} as blue dots.

Also plotted is an orange curve which describes the relationship $N^*(N)$ expected under an isotropic correlation model when the free parameters, $N^*(14)=1.96$ and $\rho=47.25\%$, are fixed by the ``final'' datum, which is the portfolio in which all assets are present. It is important to note that this curve is entirely determined by these two values \textit{and} it must pass through $(1,1)$. It is not a ``curve fit,'' adapted to the trend within the data, but an extrapolation from the final datum. As such, by eye, it appears to describe the data very well, and this statement will be made quantitative below. 

Finally, the asymptotic limiting portfolio trend $N^*(N)\approx N/K$ for an homogenously weighted $K$ factor model is plotted as a green line. This is also required to pass through the final datum and is an extrapolation from that value. The imputed value is $K=7$ and it it can be seen to be a poor description of the data, qualitatively. This is a ``large'' portfolio limiting behaviour, so it is not entirely surprising that it fails to capture the measurements made as the maximum portfolio size of fourteen could hardly be described as large.
\subsection{Hypothesis Tests for Isotropic Correlation Versus Linear Factor Models}
A summary statistic that may be computed from much of the collected data is needed to make the analysis of the fit of the various models for $N^*(N)$ quantitative. As the degrees of freedom in the portfolios formed from just one asset must be one, regardless of which of the fourteen cryptocurrencies this datum is computed from, this lower end point of the data should be excluded from the analysis. It also makes sense to excluded the upper end point as that is the value used to extrapolate the $N^*(N)$ curve from under the isotropic correlation model.

The number of ways of picking $N$ assets from $N_\textrm{max}$ is given by the binomial coefficient $N_\mathrm{max}\choose N$ and is 91 for a portfolios of two (or twelve) assets and fourteen for a portfolio of thirteen assets. In between these extrema it becomes as large as 3,432. Therefore, it is possible to summarize this data via its sample mean and standard error in that mean and to then compare this data to the extrapolation from an isotropic correlation model computed for the full portfolio (the ``final datum'') via a $\chi^2$ statistic. The usual assumption of normality in the residuals for samples over thirty in size is made and is valid for all except the portfolios of size thirteen assets. A similar statistic can be computed for fitted curves, with the degrees of freedom for the $\chi^2$ statistic reduced by the number of free parameters in the fit. The data for portfolios of one asset cannot be included in these statistics as the value of $N^*$\linebreak
\begin{landscape}
\begin{table}[p]
\begin{tabular}{rrrrrrrrrrrrr}
\multirow{2}{*}{\textbf{Assets}} & \multicolumn{4}{c}{\textbf{Sample}} & \multicolumn{4}{c}{\textbf{Isotropic Correlation}} & \multicolumn{4}{c}{\textbf{Linear Factor Model}} \\
 & \textbf{Mean} & \textbf{St.Dev.} & \textbf{Count} & \textbf{Std.Err.} & \textbf{Model} & \textbf{Error} & \textbf{$Z$ Score} & \textbf{Chi Sq.} & \textbf{Model} & \textbf{Error} & \textbf{$Z$ Score} & \textbf{Chi Sq.} \\
\hline
1 & 1.000 & 0.000 & 75 & 0.000 & 1.000 & 0.000 &  &  & 0.708 & 0.292 &  &  \\
2 & 1.306 & 0.143 & 71 & 0.017 & 1.358 & -0.052 & -3.085 & 9.515 & 1.096 & 0.210 & 12.391 & 153.542 \\
3 & 1.506 & 0.208 & 77 & 0.024 & 1.542 & -0.037 & -1.549 & 2.400 & 1.342 & 0.164 & 6.909 & 47.735 \\
4 & 1.642 & 0.239 & 64 & 0.030 & 1.655 & -0.013 & -0.421 & 0.177 & 1.510 & 0.132 & 4.399 & 19.354 \\
5 & 1.778 & 0.221 & 70 & 0.026 & 1.730 & 0.048 & 1.809 & 3.273 & 1.634 & 0.144 & 5.450 & 29.699 \\
6 & 1.777 & 0.222 & 71 & 0.026 & 1.784 & -0.007 & -0.264 & 0.070 & 1.728 & 0.050 & 1.879 & 3.530 \\
7 & 1.777 & 0.194 & 82 & 0.021 & 1.825 & -0.048 & -2.256 & 5.089 & 1.802 & -0.025 & -1.168 & 1.365 \\
8 & 1.853 & 0.193 & 50 & 0.027 & 1.857 & -0.004 & -0.141 & 0.020 & 1.862 & -0.009 & -0.312 & 0.097 \\
9 & 1.880 & 0.160 & 65 & 0.020 & 1.883 & -0.003 & -0.171 & 0.029 & 1.911 & -0.032 & -1.610 & 2.591 \\
10 & 1.902 & 0.135 & 82 & 0.015 & 1.904 & -0.002 & -0.111 & 0.012 & 1.953 & -0.051 & -3.401 & 11.569 \\
11 & 1.928 & 0.098 & 70 & 0.012 & 1.921 & 0.007 & 0.587 & 0.344 & 1.988 & -0.060 & -5.092 & 25.933 \\
12 & 1.937 & 0.096 & 71 & 0.011 & 1.936 & 0.001 & 0.075 & 0.006 & 2.019 & -0.081 & -7.130 & 50.844 \\
13 & 1.951 & 0.057 & 83 & 0.006 & 1.949 & 0.002 & 0.330 & 0.109 & 2.045 & -0.094 & -15.048 & 226.454 \\
\hline\hline
\end{tabular}

\vspace{0.5em}
\caption{Mean and standard deviation of the measured effective degrees of freedom by portfolio size for randomly selected portfolios within the Robinhood universe of tradable cryptocurrencies. These are compared to the values imputed from the final datum (the portfolio with all assets, not included in this table) for an isotropic correlation model and for the best fitting linear factor model. Data is from Robinhood from the end of July, 2024 to date.}
\tablab{meta}
\end{table}
\end{landscape}
\noindent is always unity for a portfolio of this size; the data for portfolio's of all assets is excluded from the analysis as these are the portfolios from which $N^*$ and $\rho$ are estimated and so the data must fit this final datum with zero error for the isotropic correlation model and including it would bias the $\chi^2$ statistic downwards.

The individual data are summarized in \tabref{meta} and illustrated in \figref{chisq}. A non-linear least squares procedure is used to fit the curve generated by a linear factor model to the observed data.\footnote{The \texttt{fit\_curve} function in the package \texttt{scipy.optimize} is used\cite{scipy2020}.} In \eqnref{factornstarn}, the terms $\overline{b^2}$, $\overline{\boldsymbol{b}}^T\overline{\boldsymbol{b}}$ and $\overline{s^2}$ are treated as \textit{holistically} as independent, non-negative, coefficients in the regression, rather than estimating the underlying matrix elements they are summaries of, and the estimates of them are $0.000\pm0.165$, $7.24\pm1.80\times10^7$, and $17.58\pm4.36\times10^7$, respectively. On the basis of these estimates alone, the model is poorly constrained by the data and inspection of the figure shows that the curve defined by them misses the means of the measurements almost everywhere. Mostly, it does not lie within the sampling error defined by the standard errors of the means. In contrast, the isotropic correlation model is clearly, visually, a good fit to the data and lies within the sampling error for most of the data.
\pdffigure{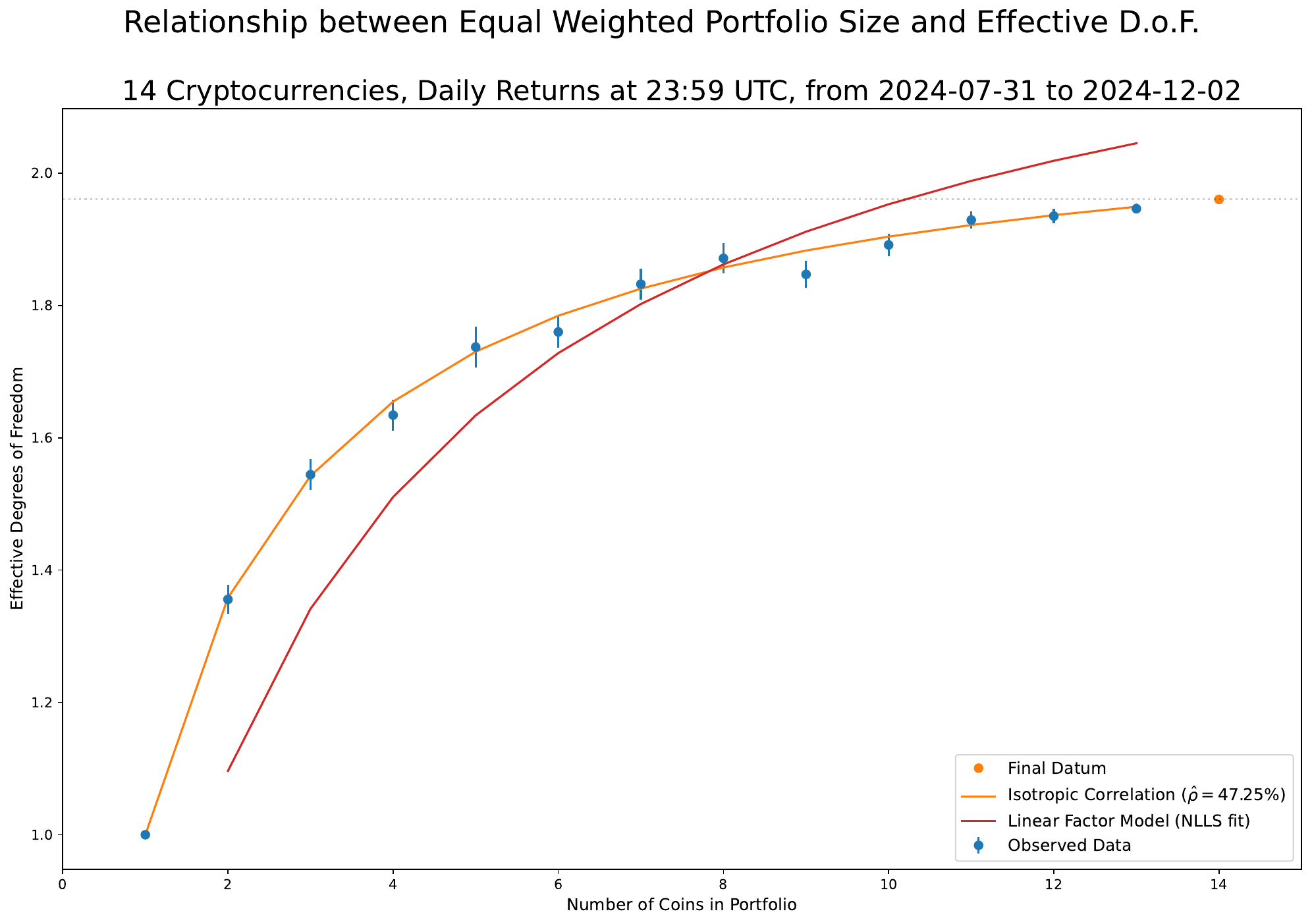}{Mean values of a random sample of values (blue dots with error bars) of $N^*(N)$ for equal weighted portfolios formed from the daily returns (UTC midnight to UTC midnight) of cryptocurrency prices collected from the \textit{Robinhood} retail brokerage and the curves expected for an isotropic correlation structure (orange curve) and the best fitting curve consistent with a linear factor model (red curve).}{0.35}{0}

The total $\chi^2$ for the linear factor model is 572.7 with 9 d.o.f., giving a vanishingly small $p$ value. In contrast, the total $\chi^2$ for the isotropic correlation model is 21.0 with 12 d.o.f.\ and a $p$ value of $0.0497$. Thus the linear factor model fails to describe the data with high confidence and the isotropic correlation model fails to describe the data with barely over 95\% confidence. The quality of the curve fit relative to the alternate isotropic model may be assessed through the $F$ statistic that is the ratio of their reduced $\chi^2$ values. This is $(572.7/9)/(21.0/12)=36.3$ which should be distributed as $F$ with 9 and 12 degrees of freedom. The associated $p$ value is vanishingly small at $4.8\times10^{-7}$, indicating that the equivalence of the models as descriptions of the data may be rejected with a high degree of confidence.
\section{Analysis of Data Collected from CoinMarketCap}
The analysis presented in the prior section appears to strongly reject the description of the collected cryptocurrency returns data as described by a standard linear factor model in favour of an isotropic correlation model. On the assumption that that description of the data is correct, it is interesting to examine how stable the parameters extracted from the model are through time --- especially since the data presented is for \textit{just} the five months prior to the date of writing.
\subsection{Data Collection}
The Robinhood data use for analysis has been collected ``live,'' by calling the API a minute before UTC midnight, and so does not extend in time before the first date in the analysis presented here. Of course, the prices for the cryptocurrencies considered do have longer history, and so that data must be acquired from another source. This data is readily available via \textit{Yahoo! Finance} and the python package \texttt{yfinance}\cite{yfinance}. The data displayed is originally sourced via CoinMarketCap\cite{coinmarketcap}, a cryptocurrency data vendor.

Data capture is straightforward using the provided API's, but the universe of cryptocurrencies is reduced as follows:
\begin{enumerate}[label=(\roman*)]
    \item\textit{Compound} (COMP-USD) is removed as the data series available terminates prior to the current year;
    \item\textit{Shiba Inu} (SHIB-USD) has very low nominal values quoted which causes numerical precision issues; and,
    \item\textit{Uniswap} (UNI-USD) exhibits an abrupt change in value by many orders of mangnitude within the middle of the data.
\end{enumerate}
\noindent Within the remaining data set \textit{Aave} (AAVE-USD) and \textit{Avalanche} (AVAX-USD) were not in existence at the beginning of the data set. Finally, in all cases, the first year of the data is arbitrarily removed to exclude high-volatility associated with initial issuance.
\subsection{Stability of Effective Degrees of Freedom Through Time}
After the data is extracted the same experiment is performed, with the additional step that the data is also stratified by year. The results of the analysis, the set of asset and portfolio variances and the values $\{(N^*_y,\rho_y)\}$ indexed by year $y\in[2018,2024]$, are presented in \tabref{variances}. The time-series of $N^*_y$ is plotted in \figref{byyear}.

For first six years studied,\footnote{i.e. Excluding the overlap with the Robinhood data.} the mean value of $N^*_y$ is 1.85 with a standard deviation of 0.38, leading to a standard error of 0.15. The value computed for the Robinhood data lies within the 68\% confidence region defined by these values: $1.96\in[1.60,2.00]$. These separate measurements are consistent on this basis are there seems to be little reason to suspect gross non-stationarity within this small sample of data.
\begin{table}[htb]
\begin{tabular}{lrrrrrrr}
&\textbf{2018} &\textbf{2019} &\textbf{2020} &\textbf{2021} &\textbf{2022} &\textbf{2023} &\textbf{2024} \\
\hline
AAVE-USD &  &  &  & 26.34 & 40.18 & 14.93 & 24.26 \\
AVAX-USD &  &  &  & 64.15 & 32.58 & 19.36 & 23.59 \\
BCH-USD & 138.10 & 28.68 & 29.95 & 48.77 & 19.11 & 17.63 & 28.42 \\
BTC-USD & 25.60 & 12.69 & 14.22 & 17.72 & 11.06 & 5.25 & 8.04 \\
DOGE-USD & 22.74 & 11.80 & 28.86 & 486.13 & 31.67 & 10.64 & 29.57 \\
ETC-USD & 50.77 & 19.13 & 27.41 & 65.96 & 36.80 & 11.24 & 17.38 \\
ETH-USD & 48.37 & 16.92 & 24.39 & 31.36 & 20.45 & 5.98 & 11.65 \\
LINK-USD & 69.62 & 48.14 & 44.24 & 53.56 & 26.50 & 14.99 & 20.37 \\
LTC-USD & 39.79 & 23.73 & 26.03 & 37.26 & 20.25 & 11.61 & 14.48 \\
XLM-USD & 41.82 & 18.44 & 37.07 & 54.79 & 16.13 & 18.61 & 30.06 \\
XTZ-USD & 50.24 & 33.48 & 35.58 & 60.44 & 23.12 & 11.13 & 25.13 \\
\hline
Portfolio ($V_P$) & 30.91 & 10.39 & 16.65 & 31.81 & 15.64 & 6.14 & 10.15 \\
Independent ($V_I$)& 4.87 & 2.13 & 2.68 & 6.57 & 1.93 & 0.98 & 1.62 \\
\hline
Total Assets ($N_\mathrm{max}$)& 10\phantom{.00} & 10\phantom{.00} & 10\phantom{.00} & 12\phantom{.00} & 12\phantom{.00} & 12\phantom{.00} & 12\phantom{.00} \\
Effective D.o.F. ($N^*$)& 1.58 & 2.05 & 1.61 & 2.48 & 1.48 & 1.92 & 1.91 \\
Imputed Correlation ($\rho$)& 59.40 & 43.07 & 57.98 & 34.90 & 64.59 & 47.74 & 47.97 \\
\hline
\hline
\end{tabular}

\vspace{0.5em}
\caption{Individual cryptocurrency variances, actual equal-weighted portfolio variance and expected portfolio variance ``as if'' the assets were indepedent, by year, together with the effective degrees of freedom, $N^*$, and imputed common pairwise correlation, $\rho$, for the reduced sample of cryptocurrencies taken from \textit{CoinMarketCap} via \textit{Yahoo! Finance}. Data for 2024 is for the partial year to the time of writing.}\tablab{variances}
\end{table}
\pdffigure{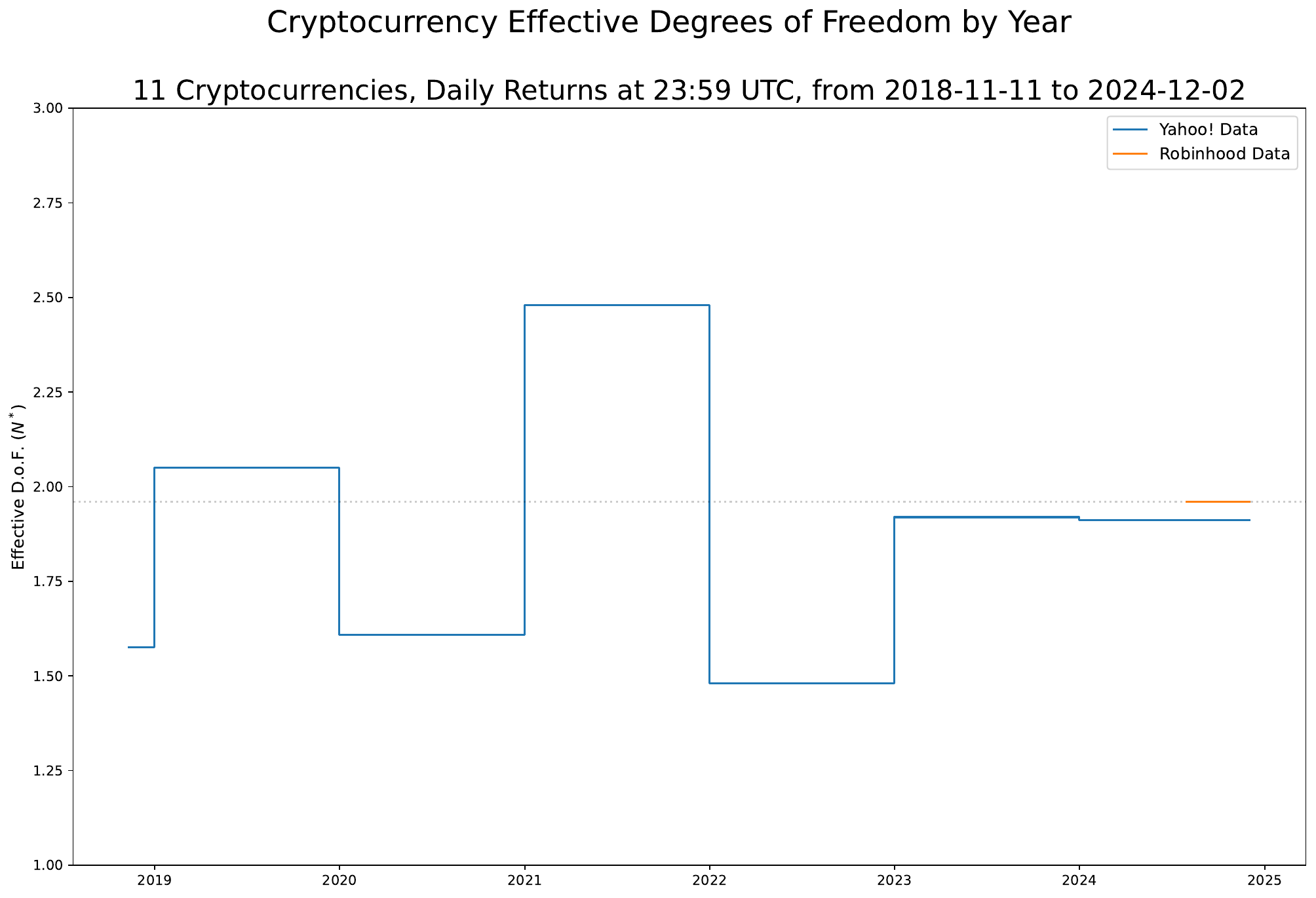}{Effective degrees of freedom from year-by-year samples of daily returns of cryptocurrencies collected from the data vendor \emph{CoinMarketCap} (blue lines). The orange line indicates the equivalent quantity extracted from the data from the retail brokerage \emph{Robinhood}.}{0.35}{0}
\section{Conclusions}
In this article, the author's prior work on isotropic correlation is applied to the emergent space of cryptocurrencies. The focus is on a very ``mainstream,'' ``retail,'' oriented set of assets: those made available to trade by clients of the popular brokerage Robinhood. The characteristic form of $N^*(N)$ is shown to be consistent with that which would be expected for a cross-section of returns described by a homoskedastic isotropic covariance structure and \textit{not} having the form that would arise from a linear factor model; not even a single factor model. Furthermore, there is evidence to support this description of the data being valid for at least the past five years.

Homoskedastic isotropic covariance is an interesting structure as, although it appears to support a ``market factor,'' there is in fact no common \textit{exogenous} driving force for returns. They just happen to be similar a lot of the time. A critical feature of this structure is that the asymptotic contribution of residual returns to a ``large'' portfolio does not vanish and it is, therefore, still possible for portfolio returns to be dominated by an individual asset. As the level of pairwise correlation is very high, at around 50\%, a mean-variance optimal asset allocator should allocate not proportional to the alpha over variance but proportional to a ``mostly centered'' $Z$-score of the alpha and inversely with respect to volatility.
\appendix
\section{Author's Statement on the Availability of Data and Code to Execute the Analysis Presented Herein}\seclab{data}
Empirical work presented here is executed in \textit{Python} code\cite{rossum1995python} using the standard ``open source'' toolkit (\textit{Pandas}\cite{mckinney2010data}, \textit{Numpy}\cite{harris2020array}, \textit{SciPy}\cite{scipy2020}) as found on Google's \textit{Colab} system\cite{googlecolab}. Analytical notebooks are archived on the author's personal \textit{GitHub} repository\cite{gillergithub} and the notebook \texttt{\smaller Longer\_Term\_Crypto\_Degrees\_of\_}\linebreak\texttt{\smaller Freedom.ipynb}, which may be found in the folder \texttt{\smaller Financial-Data-Science\-in-Python}, is used for the analysis presented herein. This code base is under development, but the version control system presented by the \textit{GitHub} website permits the specific version in use to generate the figures and tables incorporated in this document to be extracted by users. Data from the \emph{Robinhood} retail brokerage was collected via custom written code to access their provided \textsc{API} which is documented online\cite{robinhood2024api}. Proprietary code is used which is part of a trading system operated privately by the author and which is not available in the above repository. The data itself is written to a publicly available file available online\cite{giller2024cdata}. Data from \emph{CoinMarketCap} is extracted extracted programatically via the \texttt{\smaller yfinance} package\cite{yfinance} from data made available to the general public by \textit{Yahoo! Inc.}. Code to extract this data is included in the notebook given above. In both cases data updates frequently and is believed to be reliable as of the date of writing (November, 2024).
%
%
\bibliographystyle{plain}
\bibliography{citation}
\end{document}